\documentclass[manuscript]{aastex}
\slugcomment{accepted by {\it The Astrophysical Journal}, 3 April 2009}

\shorttitle{The He~I 10830~\AA| Line in HdC Stars}
\shortauthors{Geballe, Rao, \& Clayton}

\begin{document}

\title{Do Hydrogen-deficient Carbon Stars have Winds?}

\author{T. R. Geballe}
\affil{Gemini Observatory, 670 N. A'ohoku Place, 
Hilo, HI 96720; tgeballe@gemini.edu}

\author{N. Kameswara Rao} 

\affil{Indian Institute for Astrophysics, Bangalore 56034, India and W. 
J. McDonald Observatory, University of Texas at Austin, 1 University 
Station, C1400, Austin, TX 78712-0259; nkrao@iiap.res.in}

\author{Geoffrey C. Clayton}
\affil{Department of Physics and Astronomy, Louisiana State University,
Baton Rouge, LA 70803; gclayton@fenway.phys.lsu.edu}

\begin{abstract}

We present high resolution spectra of the five known hydrogen-deficient 
carbon (HdC) stars in the vicinity of the 10830~\AA\ line of neutral 
helium. In R Coronae Borealis (RCB) stars the He~I line is known to be 
strong and broad, often with a P~Cygni profile, and must be formed in 
the powerful winds of those stars. RCB stars have similar chemical 
abundances as HdC stars and also share greatly enhanced $^{18}$O 
abundances with them, indicating a common origin for these two classes 
of stars, which has been suggested to be white dwarf mergers. A narrow 
He~I absorption line may be present in the hotter HdC stars, but no line 
is seen in the cooler stars, and no evidence for a wind is found in any 
of them. The presence of wind lines in the RCB stars is strongly 
correlated with dust formation episodes so the absence of wind lines in 
the HdC stars, which do not make dust, is as expected. 

\end{abstract}


\section{Introduction}

The hydrogen-deficient carbon (HdC) stars are a very small class of 
carbon stars with very weak (or absent) hydrogen lines. Chemically they 
are completely distinct from the more common R- and N-type carbon stars. 
However, their abundances closely resemble a rare but better observed 
group of peculiar carbon stars, the R Coronae Borealis (RCB) stars 
\citep{war67,lam86}. RCB stars vary enormously in visual brightness, in 
a characteristic manner, but at irregular time intervals, due to their 
intermittent episodes of dust formation \citep{cla96}. In contrast, HdC 
stars have only weak brightness variations due to pulsations 
\citep{kil88, law97}. None of the HdC stars has shown any evidence for 
dust formation. As described in \citet{bru98}, RCB and HdC stars are 
spectroscopically similar, having the same abundances, effective 
temperatures, and absolute luminosities. However, the HdC stars lack the 
large visible brightness variations, IR excesses and emission lines seen 
in the RCB stars, which are associated with mass-loss and dust 
formation. Only five HdC stars are known \citep{war67}, but as they are 
bright stars (all are in the HD catalogue), it is likely that there are 
many more in the Galaxy \citep{war67,dri86}. The latter is also true of 
the RCB stars, of which only about 50 are known in the Galaxy 
\citep{cla96,zan05,tis08}.

The origin of RCB stars has long been debated. The leading candidates 
appear to be (1) the merger of a C/O white dwarf and a He white dwarf, 
in which the merged object undergoes an episode of helium-burning that 
creates an extended atmosphere around the degenerate core (the so-called 
double-degenerate (DD) scenario \citep{ibe84, web84}, and (2) a late or 
very late He-shell flash that blows up an isolated incipient white dwarf 
into a ``born-again" cool supergiant \citep{ibe96}.

The similar chemical abundances of HdC and RCB stars have long suggested 
that the two classes of stars are evolutionarily linked. Stronger 
evidence for a link was established recently when \citet{cla05, cla07} 
discovered that both HdC stars and RCB stars have huge overabundances of 
$^{18}$O, with $^{16}$O/$^{18}$O close to and in some cases less than 
unity (the solar and interstellar ratios are $\sim$500). \citet{gar09} 
have recently confirmed the low values of $^{16}$O/$^{18}$O in several 
of the stars observed by \citet{cla07}. Such values are not approached 
in any other class of stars. For several reasons, including the observed 
low $^{16}$O/$^{18}$O ratios, the late He-shell flash scenario seems 
unlikely to be the origin of RCB stars, and thus also unlikely to be the 
origin of the HdC stars \citep{cla07,gar09}. Therefore it is important 
to explore the suggestion that these stars are created in the mergers of 
white dwarf binaries. The initial modeling results are promising; 
conditions where material is accreting onto the more massive C/O white 
dwarf appear to be favorable for the production of copious amounts of 
$^{18}$O \citep{cla07}.

If the double degenerate merger hypothesis is correct, then the question 
arises whether the HdC and RCB stars are formed by different types of 
merger events or represent an evolutionary sequence.  Interestingly, the 
HdC stars all have lower values of $^{16}$O/$^{18}$O (i.e. more enhanced 
$^{18}$O) than RCB stars observed to date \citep{cla07}. One might 
envision an evolutionary sequence in which the RCB stars represent an 
earlier and more turbulent post-merger phase, with less $^{18}$O mixed 
to the photosphere and strong stellar winds, while HdC stars are a later 
phase when the winds have died down and mixing of $^{18}$O into the 
photosphere is complete. On the other hand it may be that the details of 
the merger process itself and/or the precise white dwarf masses create 
different degrees of $^{18}$O enhancement and the different types of 
wind environments. It is also possible that HdC stars are, in fact, RCB 
stars where the dust formation process has turned off temporarily. Some 
RCB stars show very little dust formation; e.g., XX Cam has shown only 
one small dust formation event in the last 60 years \citep{yui48, rao80}.

\section{The He 10830~\AA\ line as a probe of the wind}

\citet{cla03} observed the the He~I line at 10830~\AA\ in ten RCB stars. 
This line corresponds to the allowed 2$^3$P$_{2,1,0}$--2$^3$S$_{1}$ 
transition and has two closely spaced components at 1.0833217~$\mu$m and 
1.0833306~~$\mu$m and a third component at 1.0832057~$\mu$m (all three 
wavelengths are {\it in vacuo}) which is generally much weaker than 
either of the other components. Thus in most astronomical cases the line 
is effectively a doublet. The 2$^3$S level is metastable and cannot 
radiate to ground or be radiatively excited from the singlet (1S) ground 
state. The n=2 levels can be populated from below by collisional 
excitation (collision energies $>$20~eV for the 2$^3$S level) or from 
above by radiative decay such as would occur following recombination of 
He~II. Ionization of He~I to He~II requires either collisions or photons 
of energy $>$24~eV; the latter seems unlikely in view of the lack of 
ultraviolet radiation in the line-forming regions of RCB and HdC stars.  
Collisions with sufficient energy could be present in shocks or where 
the gas in the wind is being accelerated by dust grains.

\citet{cla03} have shown that most, if not all, RCB stars possess strong 
winds. On the other hand, the existence of winds in HdC stars has not 
been stringently tested. All that is known is that the dust clouds 
responsible for the huge light variations in RCB stars, which first form 
and then dissipate, are not present in HdC stars.  In RCB stars the He~I 
absorption lines are usually broad and blueshifted \citep{cla03}, 
implying hot expanding gas with velocities of several hundred 
km~s$^{-1}$, far greater than the escape velocities. From optical 
spectroscopy \citet{rao06} have shown that in R~CrB itself the wind 
starts in the photosphere and is heated and accelerated as it moves 
outward. The winds observed in the RCB stars are strongly correlated 
with dust formation episodes that produce large declines in visible 
brightness \citep[][and in preparation]{cla03}. During such an episode 
the He~I absorption is sometimes accompanied by a red-shifted P~Cygni 
emission feature. However, the wind line weakens or disappears within a 
few days or weeks of the ends of the decline and the return of the star 
to maximum light (Clayton et al., in preparation).

Because the He 10830~\AA\ line can reveal and characterize the winds in 
carbon-rich stars, we have used the Gemini South telescope to 
obtain high resolution spectra in the vicinity of this line in the five 
known HdC stars.  The spectra allow a direct comparison between the 
winds of the two classes of objects and thus directly address the 
relationship between the classes. A second motivation for this study is 
to investigate the nature of the wind driver. It has been suggested that 
the winds in RCB stars are dust driven \citep{cla92}. If there also 
are winds in their dustless cousins, the HdC stars, then another driving 
mechanism must exist.

\begin{deluxetable}{ccccccc} 
\tablecaption{Observing Log
\label{tbl-1}}
\tablewidth{0pt}
\tablehead{ 
\colhead{Name} & \colhead{Type} & \colhead{T$_{eff}$(K)} & 
\colhead{$V_{hel}$(km s$^{-1}$)$^{a}$} & \colhead{UT Date} & 
\colhead{Telescope} & \colhead{Calib. star}
}
\startdata
HD 137613 & HdC & 5400$^{b}$ & +56 & 20080204 & Gemini S & HIP 74493 \\
HD 148839 & HdC & 6500$^{c}$ & -24 & 20040207 & Gemini S & HIP 81873 \\
HD 173409 & HdC & 6100$^{b}$ & -69 & 20080404 & Gemini S & HIP 93667 \\
HD 175893 & HdC & 5500$^{b}$ & +44 & 20080404 & Gemini S & HIP 93367 \\
HD 182040 & HdC & 5600$^{b}$ & -46 & 20080406 & Gemini S & HIP 98953 \\
FQ Aqr    & EHe & & +16 & 20080411 & Gemini S & HIP 102631 \\
BD +23$\arcdeg$ 2998 & R2 & & -32 & 20080629 & UKIRT & HIP 79332 \\
TYC 1021 & C & & +11 & 20080629 & UKIRT & HIP 79332 \\
BD +17$\arcdeg$ 3325 & R0 & & -49 & 20080629 & UKIRT & HIP 79332 \\
XX Cam & RCB & 7250$^{d}$ & +12 & 20090108 & UKIRT & HIP 16599 \\
\enddata
\tablenotetext{a}{Values derived from these spectra, except for FQ Aqr}
\tablenotetext{b}{\citet{asp97}}
\tablenotetext{c}{\citet{law90}}
\tablenotetext{d}{\citet{asp00}}
\end{deluxetable}
\section{Observations and Data Reduction}

Spectra in the vicinity of the He I 10830~\AA\ line were obtained for 
the five known HdC stars on 2008 February 4 and 7 and April 4 and 6 at 
the Gemini South telescope, using the echelle spectrograph Phoenix, as 
part of Gemini program GS-2008A-Q-17. A log of these and additional 
observations is given in Table~1. The 0.34\arcsec\ slit was used and 
provided a resolving power of $\sim$~50,000. Each star was positioned 
alternately at two locations along the slit.  Total integration times 
were typically a few minutes. The extreme helium (EHe) star, FQ Aqr, 
which is expected to have a prominent 10830~\AA\ line, was also observed 
at Gemini South on 2008 April 11. EHe stars, which are low mass A or B 
supergiants with strong helium lines and weak or absent hydrogen lines, 
appear to be higher temperature counterparts of RCB and HdC stars 
\citep{pan01, jef08}. The wind characteristics of the EHe stars are not 
known.

Spectra of bright main sequence stars of spectral types A0--A2 were 
obtained near-simultaneously to and at similar airmasses as those of the 
HdC and EHe stars in order to remove telluric lines from the HdC and EHe 
stellar spectra and to provide wavelength calibration via the same 
telluric lines. The wavelength calibration is accurate to better than 
3~km~s$^{-1}$. A few of the A stars possessed weak 10830~\AA\ lines, 
which were artificially removed by interpolation prior to ratioing.

To obtain further comparisons, we used the United Kingdom Infrared 
Telescope (UKIRT) and its facility echelle spectrograph, CGS4 to obtain 
spectra of three normal carbon stars, TYC~1021 and 
BD~+17$\arcdeg$~3325, and BD~+23$\arcdeg$~2998, that are not expected 
to have winds and for which the He line should be weak or altogether 
absent. These stars were observed on 2008 June 29, as part of the UKIRT 
Service Program, with CGS4's narrow slit, which provided a resolving 
power of $\sim$35,000. A further UKIRT Service observation of the 
inactive RCB star XX~Cam was obtained on 2009 January 8. All of the 
UKIRT spectra were reduced in the manner described above.

\begin{figure}
\begin{center}
\includegraphics[width=12cm]{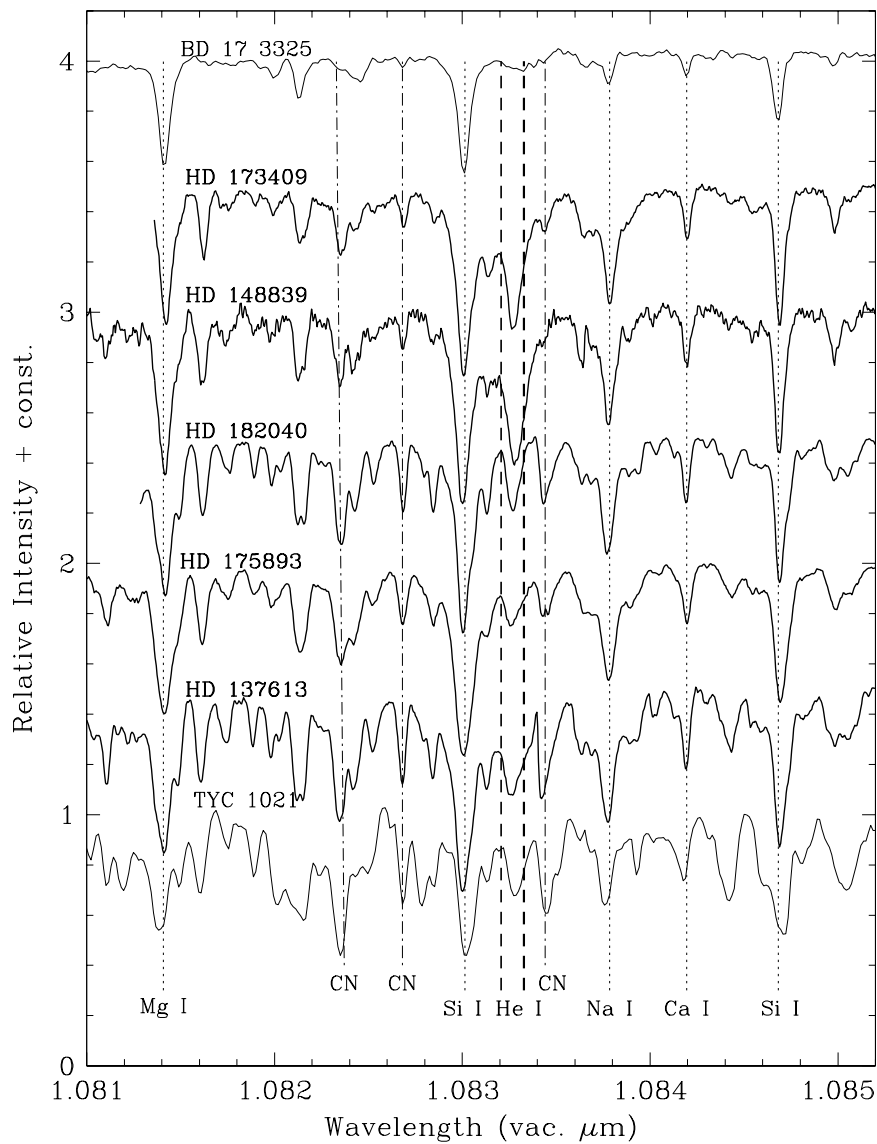}
\caption{1.081--1.085~$\mu$m spectra of the five HdC stars and 
two normal carbon stars (at top and bottom), offset vertically in steps 
of 0.5. The spectra are arranged in rough order of $T$$_{eff}$, with 
hotter stars at the top. Some of the stronger lines are identified at 
bottom, and the wavelengths of the He~I ``doublet" are also shown.}
\end{center}
\end{figure}

\begin{figure}
\begin{center}
\includegraphics[width=12cm]{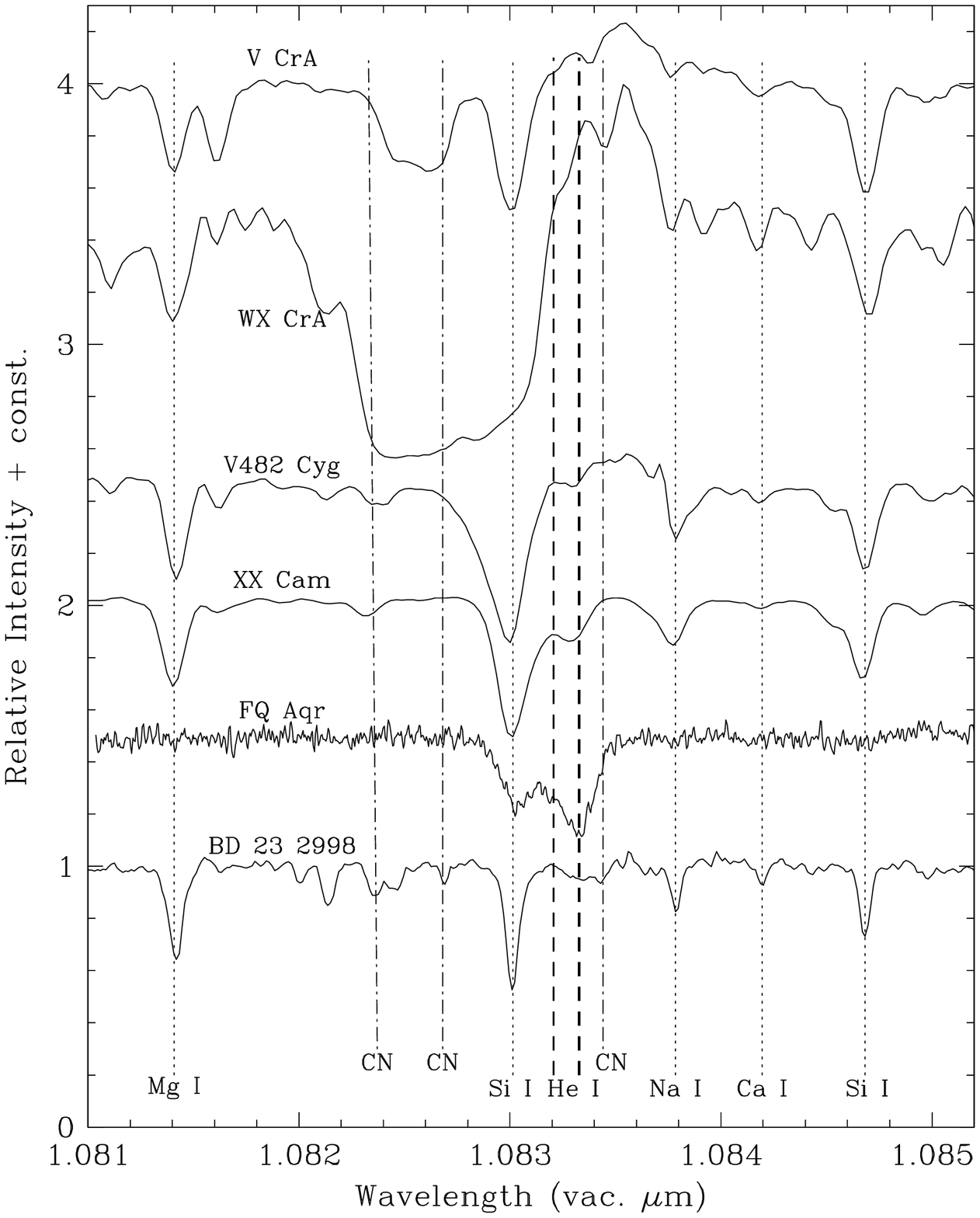}
\caption{1.081--1.085~$\mu$m spectra of the active RCB stars V 
CrA, WX CrA and V482~Cyg \citep[from][]{cla03}, the inactive RCB star 
XX~Cam, the EHe star FQ Aqr, and the hot carbon star BD~+23$\arcdeg$~2998 
offset vertically in steps of 0.5 or 1.0 (for WX CrA).}
\end{center}
\end{figure}

\section{Results}

Figure 1 contains the spectra of the five HdC stars along with the 
comparison hot carbon star BD~+17$\arcdeg$~3325 (top) and cool carbon 
star TYC~1021 (bottom). The HdC stars in Fig.~1 are arranged in order of 
effective temperature (see Table~1 for values). For most of these stars 
several wide-ranging values of $T_{eff}$ exist in the literature and 
thus the ordering may not be exact.

The wavelengths of several atomic lines, including the He~I doublet, are 
identified in both figures, as well as some of the numerous lines of CN. 
The wavelengths and identifications are from \citet{hir74}, 
\citet{wal93}, and \citet{hin95}. With the exception of FQ Aqr, the 
spectra have been shifted in wavelength to provide best overall matches 
to the atomic lines. The derived heliocentric radial velocities for the 
newly observed stars are listed in Table~1. Small deviations are present 
from line to line, but the overall accuracy of the fit is about 
3~km~s$^{-1}$. The radial velocities of the HdC stars are systematically 
about 12~km~s$^{-1}$ more negative than those in \citet{law97}.  
However, measurements of HD~137613 and HD~148839 at other wavelengths 
yield velocities within 2~km~s$^{-1}$ of the values in the Table (N. K. 
Rao, unpublished data), and thus we are confident of our values.

Figure~2 shows the spectra of the ``normal" RCB stars, V~CrA, WX~CrA, 
and V482~Cyg from \citet{cla03}, the currently inactive RCB star XX~Cam, 
the EHe star (FQ~Aqr), and another comparison carbon star, 
BD~+23$\arcdeg$~2998. V~CrA and WX~CrA show broad and complex P~Cygni 
He I profiles typical of the winds seen in RCB stars during declines. 
RCB stars that have gone a very long time without a decline at the time 
of the He~I observation show weaker He~I wind profiles (e.g., V482~Cyg) 
and in the case of XX~Cam, a spectrum very similar to the HdC stars with 
no evidence for the presence of a wind.

The spectrum of each of the three cooler HdC stars (HD~137613, 
HD~175893, and HD~182040) in Fig.~1 is a close match to the spectrum of 
the cool carbon star TYC~1021, not only in terms of the lines present 
but also in terms of the relative strengths of the lines. The spectra of 
the two hotter HdC stars (HD~173409 and HD~148839) appear to be 
intermediate between TYC~1021 and the two hotter carbon stars, 
BD~+17$\arcdeg$~3325 (Fig.~1) and BD~+23$\arcdeg$~2998 (Fig.~2). All of 
these spectra contrast strongly with that of the EHe star, FQ~Aqr 
(Fig.~2). The broad and blended absorptions in that star must both be 
due to He~I. The shorter wavelength absorption component centered at 
1.08301~$\mu$m corresponds in wavelength to a Si~I line, but the other 
Si~I line in the spectrum is absent. The longer wavelength component is 
consistent with the stellar radial velocity determined by \citet{pan01}. 
FQ Aqr thus has a broad He~I absorption centered on the stellar velocity 
and broad shell-like feature with a characteristic velocity shift of 
$\sim$90~km~s$^{-1}$.

Clearly the spectra of the HdC stars also contrast strongly with the RCB 
stars V~CrA, WX~CrA, and V482~Cyg, in that none of them possesses a 
strong and broad He~I line. They more closely resemble XX~Cam, but have 
stronger CN lines. An absorption feature centered between the 
wavelengths of the He~I doublet is present in the spectra of the HdC 
stars and in XX~Cam, but it is narrow. It also is weak in the cooler HdC 
stars. The wavelength of this feature corresponds to a line of CN; 
however CN lines in XX~Cam at other wavelengths are very weak or absent 
\citep{pan04}, which is not surprising due to its high temperature. 
Without a convincing identification and additional modelling it is 
unclear to what extent the helium line contributes to the feature.

\begin{figure}
\begin{center}
\includegraphics[width=12cm]{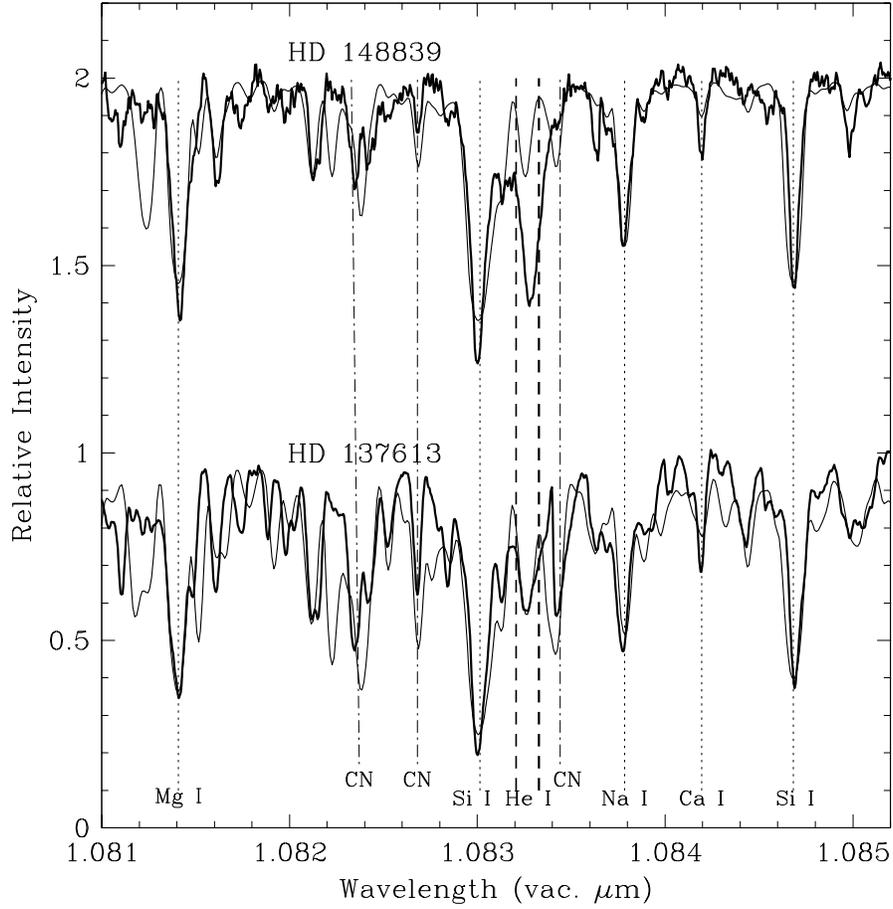}
\caption{Observed (broad lines) and synthetic (narrow lines) spectra 
of the hot HdC star HD~148839 and the cool HdC star HD~137613, 
vertically offset by 1.0 units. See text for details of models.}
\end{center}
\end{figure}

\section{Comparisons with Model Spectra}

Figure~3 shows comparisons of the spectra of the cool HdC star HD~137613 
and the hot HdC star HDC~148839 with synthetic spectra. The synthetic 
spectra, kindly produced by D. A. Garc{\'i}a-Hern{\'a}ndez, used MARCS 
model atmospheres for hydrogen-deficient stars, with scaled solar 
abundances and C/He=0.01, and included atomic lines (with the exception 
of He) from the Vienna Atomic Line Database, as well as a molecular line 
list supplied by B. Plez. The values of $T$$_{eff}$, log~$g$, and the 
abundances are the same as used by \citet{gar09}.

The agreement of the synthesized spectrum with that observed for 
HD~137613 ($T$$_{eff}$~=~5400~K, log~$g$~=~0.5) is generally good. The 
Si~I, Ca~I, Mg~I, and Na~I lines match quite well and the molecular (CN) 
features in the vicinity of the He~I line match fairly well. Only two 
prominent lines, near 1.0812~$\mu$m and 1.0822~$\mu$m do not have 
counterparts in the model, but they are far from the He~I wavelengths. 
Thus it is clear that little or no He~I absorption is present in 
HD~137613.

The same line list and parameters were used to synthesize the spectrum 
of HD~148839, using a model with $T$$_{eff}$~=~6500~K and log~$g$~=~0.5. 
The match is about as good as that for HD~137613, except near 
1.0833~$\mu$m.  Here there is an extra feature that the model does not 
reproduce. Increasing the strength of the weak CN line at this 
wavelength leads to large disagreements at other wavelengths, and indeed 
is probably unrealistic since the CN line should be slightly stronger in 
the cooler HD~137613. Similarly the strengths of weak lines of Si~I and 
Ca~I close to this wavelength cannot be increased without making the 
other lines of these species stronger than in the observed spectrum. A 
similar discrepancy between the synthetic and observed spectra is found 
for the other hot HdC star, HD~173409.

We are uncertain as to the identity of this extra feature in HD~148839. 
If it is He~I, it is blue-shifted from the dominant red member of the 
doublet by about 15~km~s$^{-1}$. Thus it would not be the typical wind 
line that is seen in RCB stars. However, if it is He~I it also could not 
be due to photospheric absorption, because of the velocity shift. If due 
to He~I, the small blue-shift would be similar to that of some narrow 
emission lines which are seen in RCB stars during declines 
\citep{ale72,cot90,rao99,sku02}.

\section{Conclusion}

The spectra presented here make it clear that just as HdC stars do not 
have the dramatic light curves and IR excesses of their cousins, the RCB 
stars, so also do they not possess winds even remotely close in strength to 
RCB stars. This result is consistent with the idea that the winds in RCB 
stars are dust-driven, which has received recent support from the 
observation that the winds weaken with increasing time after the most 
recent decline in visual brightness (G. C. Clayton et al., in 
preparation). The HdC stars also have weaker pulsations than the RCB 
stars. This difference may play a role in the absence of winds and dust 
formation in the HdC stars.

Despite the remarkable similarities in their chemical abundances and 
their even more striking similarly isotopic anomalies, suggesting a 
common origin, these two types of stars are behaving quite differently. 
We suspect that understanding the differences between RCB and HdC stars 
will require detailed studies of the white dwarf mergers that appear to 
be the best explanation for their existence.

\begin{acknowledgements}

We thank the staff of the Gemini Observatory, which is operated by the 
Association of Universities for Research in Astronomy, Inc., under a 
cooperative agreement with the NSF on behalf of the Gemini partnership: 
the National Science Foundation (United States), the Science and 
Technology Facilities Council (United Kingdom), the National Research 
Council (Canada), CONICYT (Chile), the Australian Research Council 
(Australia), Minist\'erio da Ci\'encia e Tecnologia (Brazil) and SECYT 
(Argentina). We also thank the staff of the United Kingdom Infrared 
Telescope, which is operated by the Joint Astronomy Centre on behalf of 
the Science and Technology Facilities Council of the U.K. NKR thanks A. 
Garc{\'i}a-Hern{\'a}ndez and B. Plez for assistance with synthesis of 
the spectra and D. L. Lambert for hospitality in Austin.

\end{acknowledgements}


\clearpage

\begin{thebibliography}{}

\bibitem[Alexander et al.(1972)]{ale72}Alexander, J. B., Andews, P. J., 
Catchpole, R. M., Feast, M. W., Lloyd Evans, T., Menzies, J. W., Wisse,
P. N. J., \& Wisse, M. 1972, MNRAS, 158, 305

\bibitem[Asplund et al.(1997)]{asp97}Asplund, M., Gustafsson, B., 
Kiselman, D., \& Eriksson, K. 1997, A\&A, 318, 521

\bibitem[Asplund et al.(2000)]{asp00}Asplund, M., Gustafsson, B., 
Lambert, D. L., \& Rao, N. K. 2000, A\&A, 353, 287


\bibitem[Brunner et al.(1998)]{bru98}Brunner, A.R., Clayton, G.C., \& 
Ayres, T. R. 1998, PASP, 110, 1412

\bibitem[Clayton et al.(1992)]{cla92}Clayton, G. C., Whitney, B. A., 
Stanford, S. A., \& Drilling, J. S. 1992, ApJ, 397, 652

\bibitem[Clayton(1996)]{cla96}Clayton, G.C. 1996, PASP, 108, 225

\bibitem[Clayton et al.(2003)]{cla03}Clayton, G. C., Geballe, T. R., \& 
Bianchi, L. 2003, ApJ, 595, 412

\bibitem[Clayton et al.(2005)]{cla05}Clayton, G.C., Herwig, F., Geballe, 
T. R., Asplund, M., Tenebaum, E. D., Englebracht, C. W., \& Gordon, K. 
D. 2005, ApJ, 623, L141

\bibitem[Clayton et al.(2007)]{cla07}Clayton, G. C. Geballe, T. R., 
Herwig, F., Fryer, C., \& Asplund, M. 2007, ApJ, 662, 1220

\bibitem[Cottrell et al.(1990)]{cot90}Cottrell, P. L., Lawson, W. A., \& 
Buchhorn, M. 1990, MNRAS, 244, 149

\bibitem[Drilling(1986)]{dri86}Drilling, J. S. 1986, in Hydrogen 
Deficient Stars and Related Objects, ed. K. Hunger (Dordrecht: Reidel), 
9

\bibitem[Garc{\'i}a-Hern{\'a}ndez et 
al.(2009)]{gar09}Garc{\'i}a-Hern{\'a}ndez, D. A., Hinkle, K. H., 
Lambert, D. L., \& Erikkson, K. 2009, ApJ, in press (astro-ph 0901.3667)

\bibitem[Hinkle et al.(1995)]{hin95}Hinkle, K., 
Wallace, L., \& Livingston, W. 1995, PASP, 107, 1042

\bibitem[Hirai(1974)]{hir74}Hirai, M. 1974, PASJ, 26, 163

\bibitem[Iben \& Tutukov(1984)]{ibe84}Iben, I. J. \& Tutukov, A. V. 
1984, ApJS, 54, 335

\bibitem[Iben et al.(1996)]{ibe96}Iben, I. J., Tutukov, 
A. V., \& Yungleson, L. R. 1996, ApJ, 456, 750

\bibitem[Jeffery(2008)]{jef08}Jeffery, C. S. 2008, in Hydrogen-Deficient 
Stars, K Werner and T Rausch, eds. 2008, ASP Conf Ser. 291, 53

\bibitem[Kilkenny et al.(1988)]{kil88} Kilkenny, D. Marang, 
F., \& Menzies, J. W. 1988, MNRAS, 233, 209

\bibitem[Lambert(1986)]{lam86}Lambert, D. L. 1986, in Hydrogen Deficient 
Stars and Related Objects, ed. K. Hunger (Dordrecht: Reidel), 127

\bibitem[Lawson et al.(1990)]{law90}Lawson, P. L., Cotrell, P. L., 
Kilmartin, P. M., \& Gilmore, A. C. 1990, MNRAS, 247, 91

\bibitem[Lawson \& Cottrell(1997)]{law97}Lawson, W. A. \& Cottrell, P. 
L. 1997, MNRAS, 266, 276

\bibitem[Pandey et al.(2001)]{pan01}Pandey, G., Rao, N. K., Lambert, D. 
L., Jeffery, C. S., Asplund, M. 2001, MNRAS, 324, 937

\bibitem[Pandey et al.(2004)]{pan04}Pandey, G., Lambert, D. L., Rao, N. K., 
Gustafsson, B., Ryde, N, \& Yong, D. 2004, MNRAS, 353, 143

\bibitem[Rao et al.(1980)]{rao80}Rao, N. K., Ashok, N. M., \& Kulkarni, 
P. V. 1980, J. Ap\&A, 1, 71

\bibitem[Rao et al.(1999)]{rao99}Rao, N. K., et al. 1999, MNRAS, 310, 
717

\bibitem[Rao et al.(2006)]{rao06}Rao, N. K., Lambert, D. 
L., \& Shetrone, M. D. 2006, MNRAS, 370, 941

\bibitem[Skuljan \& Cottrell(2002)]{sku02}Skuljan, Lj. \& Cottrell, P. 
L. 2002, MNRAS, 335, 1133

\bibitem[Tisserand et al.(2008)]{tis08}Tisserand, P. et al. 2008, A\&A, 
481, 673

\bibitem[Wallace  et al.(1993)]{wal93}Wallace, L., 
Hinkle, K., \& Livingston, W., 1993, An Atlas of the Photospheric 
Spectrum from 8900 to 13600 cm$^{-1}$ (7350 to 11230~\AA), National 
Solar Observatory, N.S.O. Technical Report 93-001

\bibitem[Warner(1967)]{war67}Warner, B. 1967, MNRAS, 137, 119

\bibitem[Webbink(1984)]{web84}Webbink, R. F. 1984, ApJ, 277, 355

\bibitem[Yuin(1948)]{yui48}Yuin, C. 1948, AJ, 107, 413

\bibitem[Zaniewski et al.(2008)]{zan05}Zaniewski, A., Clayton,G.C., 
Welch, D.L., Gordon, K.D., Minniti, D., \& Cook, K.H. 2005, AJ, 130, 
2293


\end{thebibliography}
\end{document}